\RequirePackage[2020-02-02]{latexrelease}
\documentclass[aps,prb, twocolumn, superscriptaddress,floatfix]{revtex4-2}%

\usepackage{amsfonts}
\usepackage{amsmath}
\usepackage{amssymb}
\usepackage{graphicx}
\usepackage{epstopdf}
\usepackage{textcomp}
\usepackage[normalem]{ulem}
\usepackage{color}
\usepackage{gensymb}
\usepackage{float}

\begin{document}

\title{Multimodal synchrotron X-ray diffraction across the superconducting transition of Sr$_{0.1}$Bi$_2$Se$_3$}

\author{M. P. Smylie}
\affiliation{Materials Science Division, Argonne National Laboratory, 9700 S. Cass Ave., Lemont, Illinois 60439, USA}
\affiliation{Department of Physics and Astronomy, Hofstra University, Hempstead, New York 11549, USA}

\author{Z. Islam}
\affiliation{Advanced Photon Source, Argonne National Laboratory, Lemont, Illinois 60439}

\author{G. D. Gu}
\author{J. A. Schneeloch}
\author{R. D. Zhong}
\affiliation{Condensed Matter Physics and Materials Science Department, Brookhaven National Laboratory, Upton, New York 11793}

\author{S. Rosenkranz}
\author{W.-K. Kwok}
\author{U. Welp}
\affiliation{Materials Science Division, Argonne National Laboratory, 9700 S. Cass Ave., Lemont, Illinois 60439, USA}

\begin{abstract}
In the doped topological insulator Sr$_x$Bi$_2$Se$_3$, a pronounced in-plane two-fold symmetry is observed in electronic properties below the superconducting transition temperature $T_c \sim 3$ K, despite the three-fold symmetry of the observed $R\bar{3}m$ space group.
The axis of two-fold symmetry is nominally pinned to one of three rotationally equivalent directions, and crystallographic strain has been proposed to be the origin of this pinning.
We carried out multimodal synchrotron diffraction and resistivity measurements down to $\sim$0.68 K and in magnetic fields up to 45 kG on a single crystal of Sr$_{0.1}$Bi$_2$Se$_3$ to probe the effect of superconductivity on the crystallograpic distortion.
Our results indicate that there is no in-plane crystallographic distortion at the level of $1\times10^{-5}$ associated with the superconducting transition.
These results further support the model that the large two-fold in-plane anisotropy of superconducting properties of Sr$_x$Bi$_2$Sr$_3$ is not structural in origin but electronic, namely it is caused by a nematic superconducting order parameter of $E_u$ symmetry.
\end{abstract}

\date{\today}

\maketitle

\section{Introduction}
Following the discovery of topological insulators \cite{Fu-TI-prediction, Zhang-Bi2Se3-predict, XiaBi2Se3-discovery, HsiehBi2Se3-discovery, Mazumder-Bi2Se3}, with an insulating gap in the bulk and a gapless conductive surface state, it was quickly realized \cite{FuKane2008, Schnyder2008, QiZhang2009, HasanKane2010, QiZhang2011, TanakaReview2012, SasakiReview2015, Hasan2015Review, MizushimaReview2016} that the superconducting version, the topological superconductor, could exist.
A topological superconductor can have a nodeless or nodal superconducting gap in the bulk \cite{SchnyderBrydon}, while simultaneously possessing gapless surface states.
These surface states may support quasiparticle excitations which are Majorana zero-modes, whose non-Abelian nature could be used to create a robust quantum computer \cite{Wilczek2009, Beenakker2009, Flensberg2012}.
Two methods have been used to generate topological superconductivity: (1) via proximity effect \cite{ProximityMourik, ProximityDas, ProximityRokhinson, ProximityBeenakker, ProximityAlbrecht} by deposition of a conventional $s$-wave superconductor onto a topological insulator surface and (2) via doping a topological insulator \cite{Erickson-SnInTe, HorCBS, Sasaki2012-CBS-SIT, Zhong-PbSnInTe-superconductivity, Polley-SnInTe-ARPES} to evoke bulk superconductivity.
The former approach has shown great promise, but definitive Majorana detection remains controversial \cite{MajoranaRetraction, CastelvecchiNatureNewsMajorana2021}.
The latter method has revealed some unexpected electronic behavior.

Doping with either Cu, Nb, or Sr \cite{HorCBS, HorNBS, LiuSBS2015} induces superconductivity in the well-known topological insulator Bi$_2$Se$_3$ \cite{Mazumder-Bi2Se3} while preserving its topological order \cite{WrayCBS-NatPhys}.
The first observation of superconductivity was reported in Cu$_x$Bi$_2$Se$_3$ \cite{HorCBS} with a $T_c$ of $\sim$3.4 K and a full superconducting gap \cite{CBS-STM}.
However, this material is not air-stable and reported superconducting volume fractions in single crystals are typically low \cite{HorCBS, CBS-Kriener-Synthesis2011, CBS-Kondo-Synthesis2013, CBS-Wang-Synthesis2016}.
Nb$_x$Bi$_2$Se$_3$ has a similar $T_c \sim$3.4 K, albeit the synthesis of this material remains challenging \cite{NBS-Kobayashi-Synthesis2017, NBS-Wang-Synthesis2020, NBS-Kamminga-Synthesis2020, NBS-Kevy-Synthesis2021, NBS-Dalgaard-Synthesis2021}.
In particular, an unresolved issue for Cu and Nb-doping is the determination of the exact location of the dopant ions in the superconductor.
Recent reports \cite{NBS-Dalgaard-Synthesis2021, CBS-Frolich-ND, Li-PRM-SBS-TEM, Yu-PRB-CBS-TEM, LinAndo-PRB-SBS-NIXSW} yield conflicting results as to whether the dopant ions are intercalated in the van der Waals gap, are incorporated in other locations in the lattice, or undergo clustering \cite{NBS-Goutam-STM}.
Quantum oscillation measurements on Nb-doped Bi$_2$Se$_3$ \cite{NBS-Lawson-dHvA} indicate multiple Fermi surface sheets, and penetration depth \cite{SmylieNBS1, SmylieNBS2} and STM measurements \cite{NBS-Goutam-STM} find a nodal superconducting gap structure.
Sr doping \cite{LiuSBS2015}, on the other hand, generates superconductivity at $T_c \sim$3.0 K, and millimeter-scale stable crystals can be grown with nearly 100\% superconducting volume fraction \cite{SBS-Shruti-Synthesis, SBS-Han-Synthesis}.
STM measurements \cite{SBS-STM} suggest a full superconducting gap.
In the Cu- and Sr-doped compounds, quantum oscillations and ARPES measurements \cite{CBS-Lawson-dHvA, Lawson-PRB-CBS-QO, CBS-Lahoud-SdH, LiuSBS2015, SBS-Almoalem-SdH} show only one cylindrical Fermi surface sheet, and for Cu$_x$Bi$_2$Se$_3$, the Fermi surface is found to undergo a Lifshitz transition from closed ellipsoidal to an open warped cylindrical Fermi surface \cite{CBS-Lahoud-SdH,CBS-Lifshitz} upon sufficient doping to elicit superconductivity.

All three compounds with Cu, Nb, and Sr doping share the same $R\bar{3}m$ trigonal structure of the parent compound Bi$_2$Se$_3$.
Therefore, the observation of a pronounced two-fold in-plane asymmetry in of the superconducting state is unexpected \cite{YonezawaReview}.
First observed in Knight shift measurements on Cu$_x$Bi$_2$Se$_3$ \cite{CBS-Knight}, this behavior was interpreted as signature of odd-parity superconductivity \cite{FuCBS-PRB2014}.
Subsequently, a pronounced two-fold basal plane symmetry was observed in magnetoresistivity, calorimetry, torque magnetometry, and upper critical field measurements \cite{CBS-Yonezawa-CT, SBS-Pan-SciRep, NBS-Asaba-PRX, SBS-KWilla-CT, SBS-Sun-CT, YonezawaReview} in the superconducting state of the three Bi$_2$Se$_3$ derived superconductors Cu$_x$Bi$_2$Se$_3$, Nb$_x$Bi$_2$Se$_3$, and Sr$_x$Bi$_2$Se$_3$ despite their three-fold symmetric crystal structure.
STM measurements \cite{CBS-STM-nematic} directly show a two-fold symmetric superconducting gap in Cu$_x$Bi$_2$Se$_3$.
A pseudo-spin triplet, nematic superconducting state with a two-component order parameter has been proposed \cite{FuCBS-PRB2014, VenderbosCBS-PRB2016} to account for the observed two-fold symmetry below $T_c$.
This state is odd-parity and has $E_u$ symmetry and allows for the possibility of an anisotropic full gap as well as a nodal gap.
Either gap structure qualifies as topological \cite{SchnyderBrydon}, and the observation of zero-bias conductivity peaks in spectroscopy measurements \cite{CBS-Sasaki-ZBCP, CBS-Kirzhner-ZBCP, NBS-Kurter-ZBCP} has been interpreted as a signature of Majorana states on the surface.
This rotational symmetry breaking state and its observed insensitivity to disorder \cite{SmylieNBS2, NagaiUnUnconventional, NewAnderson, Cavanagh-Disorder} identify the doped Bi$_2$Se$_3$ superconductors as a new type of unconventional superconductor as both the gap amplitude and phase are lower symmetry than that of the lattice.

Theoretical models predict that the nematic director is aligned with either the $a$ or the $a^*$ direction in the crystal. As each has three rotationally equivalent directions due to the three-fold symmetry in the $R\bar{3}m$ structure, the appearance of three equivalent nematic domains is expected preserving the overall symmetry of the $R\bar{3}m$ structure.
However, the majority of experimental reports show a single nematic axis accompanied by a pronounced two-fold in-plane anisotropy.
We have previously observed the nematic axis to be pinned to one in-plane $a$ axis.
For a given crystal this axis does not change upon repeated thermal cycling to room temperature.
A nematic axis along the $a$-axis is consistent with the proposed nodal $\Delta4_x$ state.
However, other groups have reported \cite{CBS-D4xD4y-Kawai, SBSCorbino} on crystals with $a^*$ pinning, consistent with the proposed anisotropic but fully gapped $\Delta4_y$ state.
One report presents angular dependent magnetotransport data on a single crystal \cite{SBS-Kostylev-domains} which showed the three rotationally equivalent nematic axis configurations simultaneously present.
Theoretical models \cite{VenderbosCBS-PRB2016, AkzyanovMBS-PRB2020, HowYip2019} predict that the superconducting order parameter couples linearly to strain fields, thereby providing a mechanism for the selection of a nematic axis through residual strains, for instance.

Alternatively, a structural transition into a two-fold symmetric state would naturally explain the observed superconducting anisotropy.
No transitions aside from superconductivity have been reported in magnetotransport or calorimetry \cite{SBS-Pan-SciRep, YonezawaReview, SBS-KWilla-CT, SmylieSciRep} nor in recent neutron diffraction measurements on Cu$_x$Bi$_2$Se$_3$ \cite{CBS-Frolich-ND}.
Room-temperature high-L reflection-geometry synchrotron measurements \cite{SmylieSciRep} show no distortions from the $R\bar{3}m$ structure.
However, Kuntsevich et al \cite{KuntsevichXRD2018, KuntsevichXRD2019} report a 0.02\% monoclinic distortion in the (2 0 5) and (1 1 15) peaks at room temperature in single-crystal Sr$_x$Bi$_2$Se$_3$.
Additionally, Cho et al \cite{NBS-Lortz} report distortion of the lattice along one direction just above $T_c$ via dilatometry measurements, suggesting that the structural distortion could be masked in other bulk probes by the superconducting transition.
While it is unlikely that the small reported distortion ($\Delta L/L \sim 10^{-7}$) could account for the large in-plane superconducting anisotropy of $\Gamma \sim $3-4 at T = 0 \cite{SBS-Pan-SciRep, SmylieSciRep}, it may nevertheless pin the nematic axis. 

Here, we present simultaneous resistivity and XRD measurements on a Sr$_{0.1}$Bi$_2$Se$_3$ crystal.
By monitoring the sample’s resistance, we measure the superconducting transition as a function of temperature and applied magnetic field while recording the in-plane Bragg reflections.
Our results indicate the absence of any in-plane crystallographic distortion at the level of $1\times10^{-5}$ associated with the superconducting transition.
These results further support the model that the large two-fold in-plane anisotropy of superconducting properties of Sr$_x$Bi$_2$Se$_3$ is not structural in origin but electronic.
Namely, it is caused by a nematic superconducting order parameter of $E_u$ symmetry.
Our multimodal measurement technique combining simultaneous sub-kelvin magnetotransport and diffraction measurements on a single crystal at a synchrotron beamline serves as a proof-of-concept experiment which may open new avenues for materials science research.

\section{Experimental Methods}

A relatively large $\sim$ 2 cm ($l$) x 0.65 cm ($w$) x 0.13 cm ($h$) rectangular platelike single crystal was cleaved from a bulk crystal, grown via melt-growth technique \cite{SBSCorbino}.
The crystal was screened for superconductivity in a custom-built SQUID magnetometer with a small conventional magnet.
Gold electrical contact strips were evaporated onto the long face of the crystal, and gold wires were then attached to the strips using silver epoxy in a conventional four-point measurement configuration, with the current flow of 0.1 mA in the $a$-$a^*$ plane.
An AMI 90/10/10 kG superconducting 3-axis vector magnet with a standard $^4$He variable temperature insert was used for magnetotransport measurements before and after the synchrotron characterization.
The vector magnet allowed for the magnetic field to be swept in the $a$-$a^*$ plane in the crystal without having to physically rotate the sample.
\begin{figure}[h]
    \includegraphics[width=1\columnwidth]{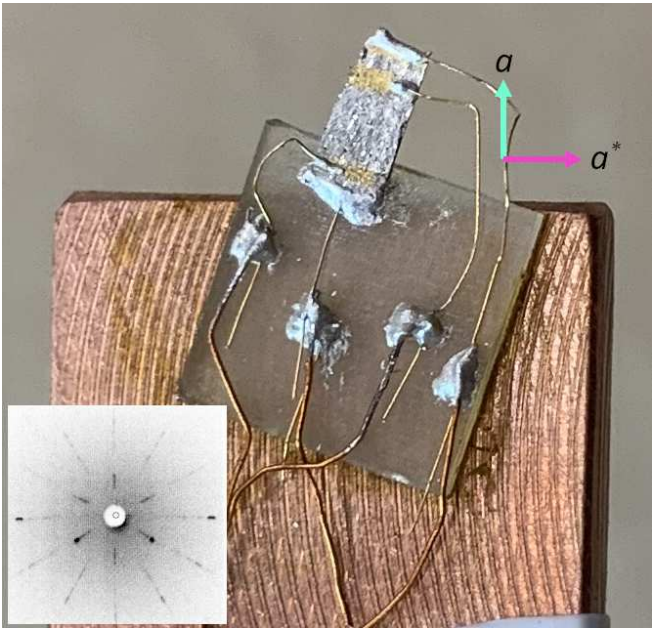}
    \caption{
Single crystal of Sr$_{0.1}$Bi$_2$Se$_3$, wired for transport measurements and mounted on a sapphire wafer on the cold stage of a $^3$He cryostat suitable for transmission XRD.
The sample is oriented such that the $a$ axis of the crystal is parallel to the long axis of the probe (and thus parallel to the applied magnetic field).
The inset shows a Laue pattern of the single crystal of Sr$_{0.1}$Bi$_2$Se$_3$, used for orienting the sample.
}
    \label{figPhoto}
\end{figure}
For the synchrotron measurements, the sample was mounted at one end onto a sapphire wafer with silver epoxy while the other end extended beyond the sapphire substrate thereby enabling unobstructed x-ray transmission (see Fig.~\ref{figPhoto}).
\begin{figure*}[!ht]
    \includegraphics[width=2\columnwidth]{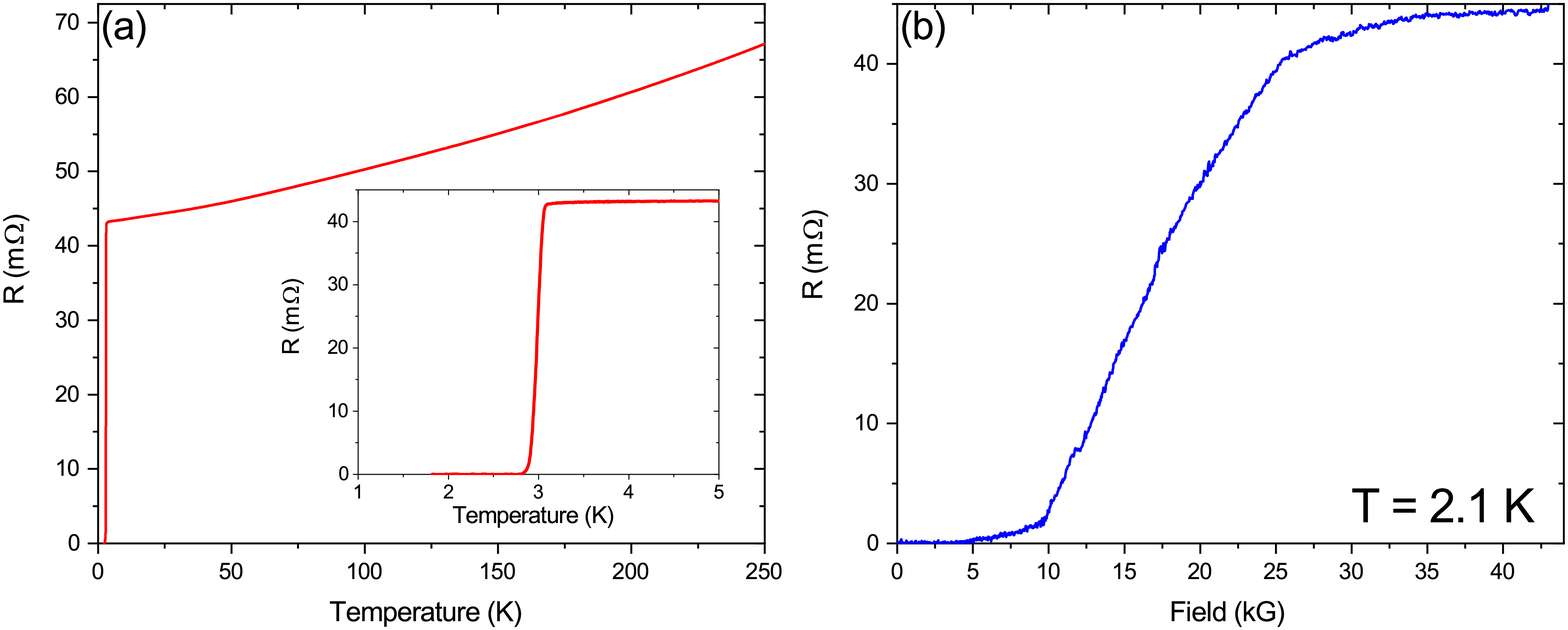}
    \caption{
(a) Resistance vs temperature of the single crystal of Sr$_{0.1}$Bi$_2$Se$_3$ selected for synchrotron measurements in zero applied field as measured in a conventional $^4$He exchange gas cryostat.
The inset shows the transition on expanded scales; $T_{c,onset} \approx$ 3.0 K.
(b) Resistance at 2.1 K as a function of magnetic field applied along the $a$ axis of the crystal, obtained at the synchrotron cryostat with the x-ray beam off. At 2.1 K, a maximum field of 45 kG is enough to drive the sample into the normal state.
}
    \label{figRTRH}
\end{figure*}
With the help of Laue pictures (Fig.~\ref{figPhoto}, inset) the sample was aligned such that the $a$-axis was oriented parallel to the cryostat axis.
X-ray measurements were performed at the 6-ID-C beamline at the Advanced Photon Source at Argonne National Laboratory with a beam energy of 19.9 keV, sufficiently high to ensure reasonable transmission.
Beam slits were positioned such that an illuminated area of approximately 300 $\mu$m x 300 $\mu$m was located between the two voltage contacts.
Transport measurements were performed in-situ in the x-ray beam in magnetic fields of up to 45 kG generated by a split-coil superconducting magnet affording wide horizontal optical access.
The transport wires were anchored at multiple places on the probe and cold-head to minimize heat loading from room temperature to the $^3$He pot.
Currents of 0.1 mA were used in all transport measurements on the $^3$He cold finger, and a ramp rate of 0.1 K/min was used at low temperature to measure all R(T) curves.
With the sample mounted as shown in Fig.~\ref{figPhoto}, the $^3$He cold finger thermometer attached to the $^3$He pot reached a base temperature of 0.68 K with a hold time of approximately 30 minutes, sufficiently long to record x-ray scans.


\section{Results and Discussion}
Fig.~\ref{figRTRH}(a) shows the temperature dependence of the resistance of the Sr$_{0.1}$Bi$_2$Se$_3$ crystal as measured in the AMI vector magnet in zero field.
We find a residual resistivity ratio of about 1.45, and a sharp superconducting transition with an onset at approximately 3.0 K and a transition width $\Delta T_c<$ 0.2 K, typical for Sr$_{0.1}$Bi$_2$Se$_3$ crystals.
Figure 2(b) shows the field dependence of the resistance of the crystal at 2.1 K with H $\parallel$ $a$ as measured in the $^3$He cold finger cryostat at the synchrotron in no beam; in 45 kG, the sample is fully in the normal state.

\begin{figure}
    \includegraphics[width=1\columnwidth]{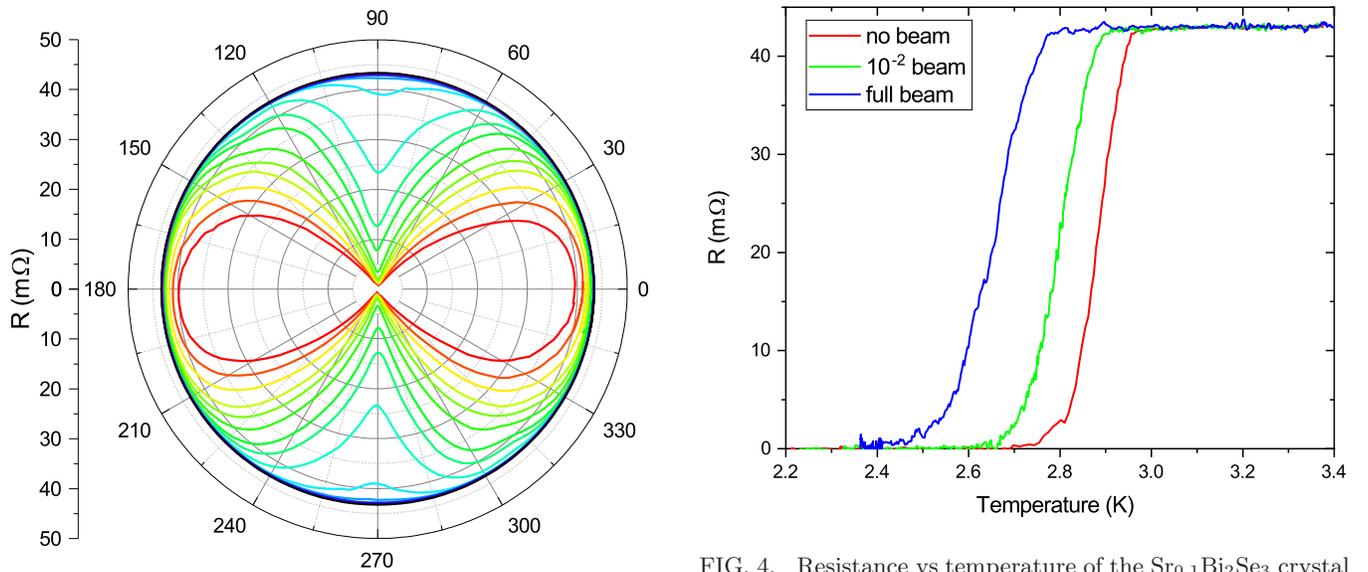}
    \caption{
R($\theta$) with H = 10 kG swept in the $a$-$a^*$ plane for the same crystal showing the two-fold axis of symmetry, with the high-$H_{c2}$ direction pinned to the $a$ axis (90$\degree$). Temperatures start at 1.8 K (innermost, red) and increase in 0.1 K steps until the normal state isotropic behavior is recovered above $T_c$ (outermost, black).
}
    \label{figRTheta}
\end{figure}

The resistance of the Sr$_{0.1}$Bi$_2$Se$_3$ crystal as a function of in-plane angle in an applied magnetic field of 10 kG in the AMI vector magnet cryostat is shown in Fig.~\ref{figRTheta} for temperatures from 1.8 K (red) to above $T_c$ (black) in 0.1 K steps.
The angular dependence of the resistance shows the characteristic two-fold anisotropy of the upper critical field $H_{c2}$ observed in the nematic superconducting state.
Directions with lower $H_{c2}$ will be resistive, whereas directions with higher $H_{c2}$ remain superconducting; as temperature increases, the anisotropy lifts.
Laue x-ray measurements (inset of Fig.~\ref{figPhoto}) on the same crystal confirm that the high-$H_{c2}$ direction (marked as 90$\degree$ in Fig.~\ref{figRTheta}) is a crystallographic $a$-axis, consistent with all previous measurements on crystals of Sr$_x$Bi$_2$Se$_3$ by our group. 

\begin{figure}
    \includegraphics[width=1\columnwidth]{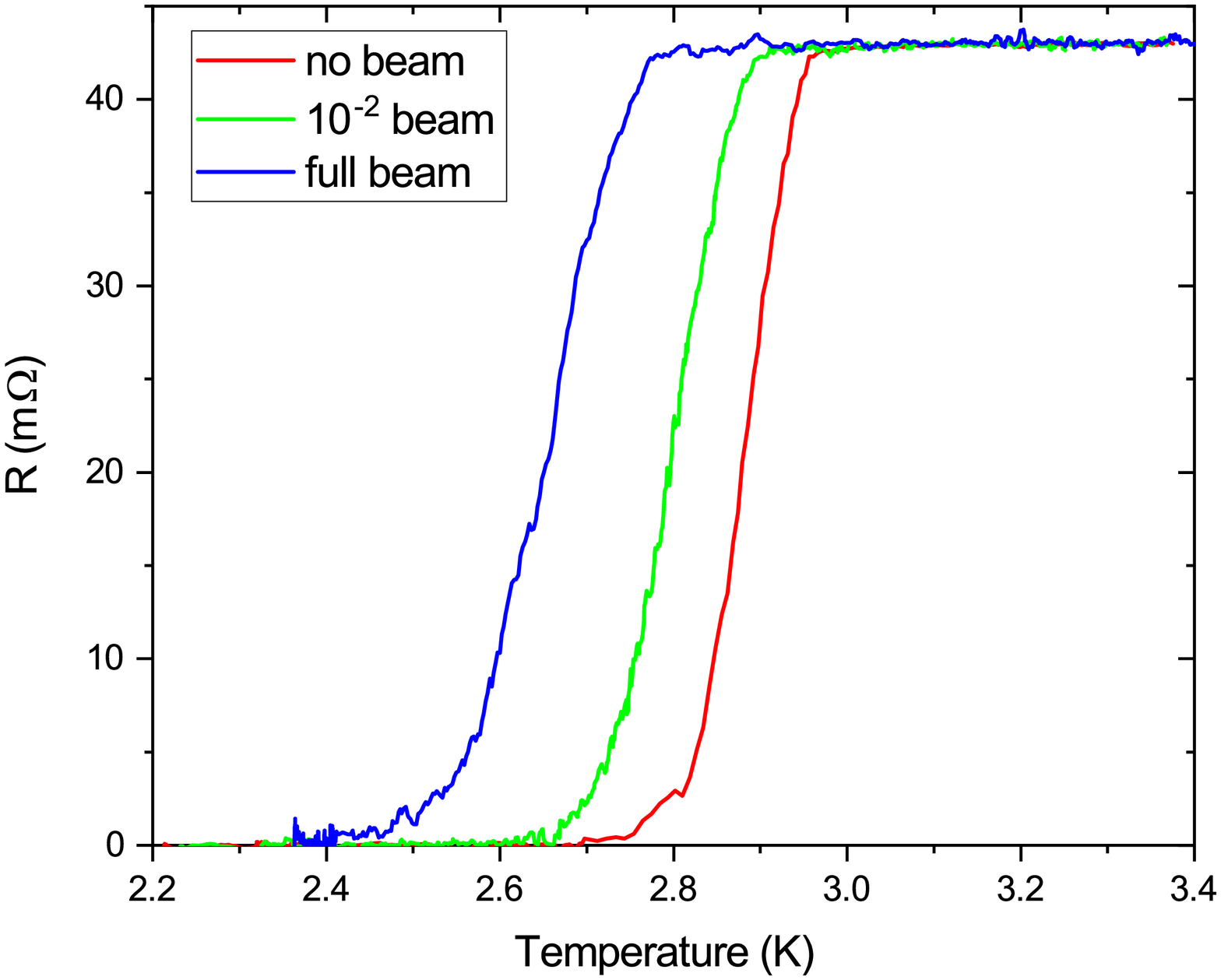}
    \caption{
Resistance vs temperature of the Sr$_{0.1}$Bi$_2$Se$_3$ crystal in-situ on the synchrotron stage demonstrating beam-induced heating.
With the full beam (10$^{11}~\gamma$/sec), the transition is only lowered by $\sim$250 mK.
}
    \label{figRTBeam}
\end{figure}

A persisting challenge in low-temperature XRD measurements is assessing the "true" sample temperature which, due to beam induced heating, may be significantly higher than the thermometer reading \cite{Islam-Rydh}.
In our multimodal set-up, we use the temperature dependence of the resistance and the location of the superconducting transition as indication of the actual sample temperature.
Fig.~\ref{figRTBeam} shows in-situ resistance vs temperature measurements under different levels of beam load.
With full beam of approximately 10$^{11}~\gamma$/sec, the transition is suppressed by only $\sim$250 mK; different beam attenuation levels were seen to shift the superconducting transition temperature between these two extrema.
This suggests that around 2 K, beam-induced heating raises the sample temperature by at most 0.25 K above the thermometry reading.

\begin{figure}
    \includegraphics[width=1\columnwidth]{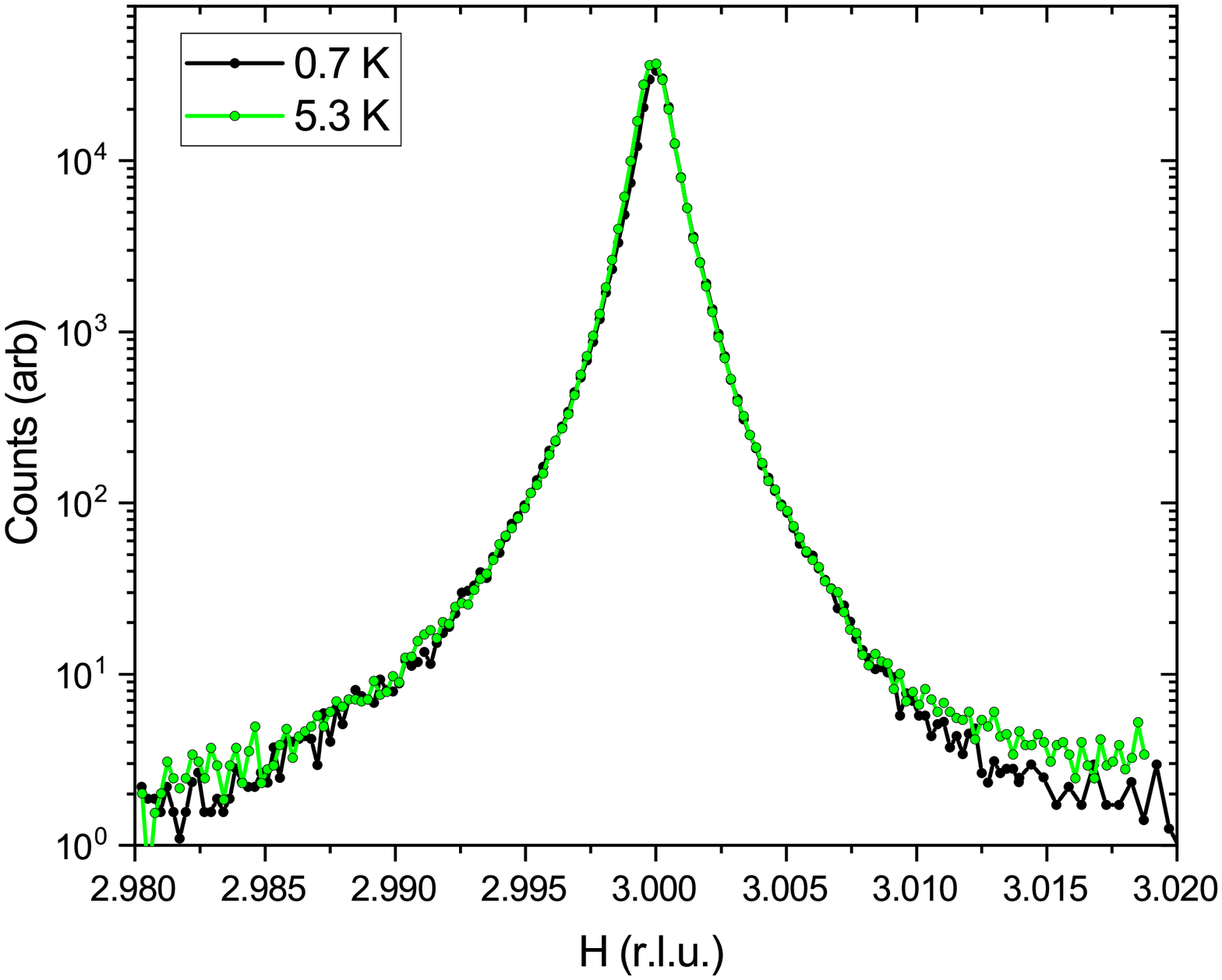}
    \caption{
(300) peak in the $R\bar{3}m$ structure deep in the superconducting state (0.69 K, black) and above $T_c$ (5.3 K, green).
There is no discernable difference, suggesting no symmetry-breaking distortion of the lattice.
}
    \label{figNoField}
\end{figure}

A monoclinic distortion, as reported by Kuntsevich $et al.$ \cite{KuntsevichXRD2018, KuntsevichXRD2019}, should manifest itself as a splitting and/or shift of high-symmetry peaks as the three rotationally symmetric $a$-directions in the nominal $R\bar{3}m$ trigonal structure would no longer be equivalent.
Here we report results on the (300) reflection.
Fig.~\ref{figNoField} shows H-scans in a narrow window centered around H = 3 recorded at base temperature of the system (0.68 K, black), and above $T_c$ (5.3 K, green).
There is no discernable difference between the two scans, strongly suggesting that there is no crystallographic transition at or close to $T_c$ which could be masked by the superconducting signal in measurement techniques such as magnetization, magnetotransport and calorimetry.
Our results are also inconsistent with any monoclinic distortion away from the nominal three-fold symmetric $R\bar{3}m$ structure.

In addition to varying the sample temperature, the application of a magnetic field allows driving the sample into the normal state while holding the temperature constant.
With our sample geometry (Fig.~\ref{figPhoto}), the field was applied along the crystallographic $a$-axis, which is the axis of high $H_{c2}$.
The sample was kept at T = 2.1 K while the field was ramped from 0 to 45 kG.
The field was ramped slowly to avoid eddy current heating, and no significant sample stage heater output variations were observed during ramping, suggesting that any field-induced heating or variations in thermometry were negligible.
Fig.~\ref{figRTRH}(b) shows the field dependence of the sample at 2.1 K with the beam off.
The onset of resistance occurs around 6 kG, which is consistent with values at 2.1 K for similar crystals.
These data demonstrate that the applied magnetic field is able to drive the sample normal while on the cold finger in the beamline cryostat.

\begin{figure*}[t]
    \includegraphics[width=2\columnwidth]{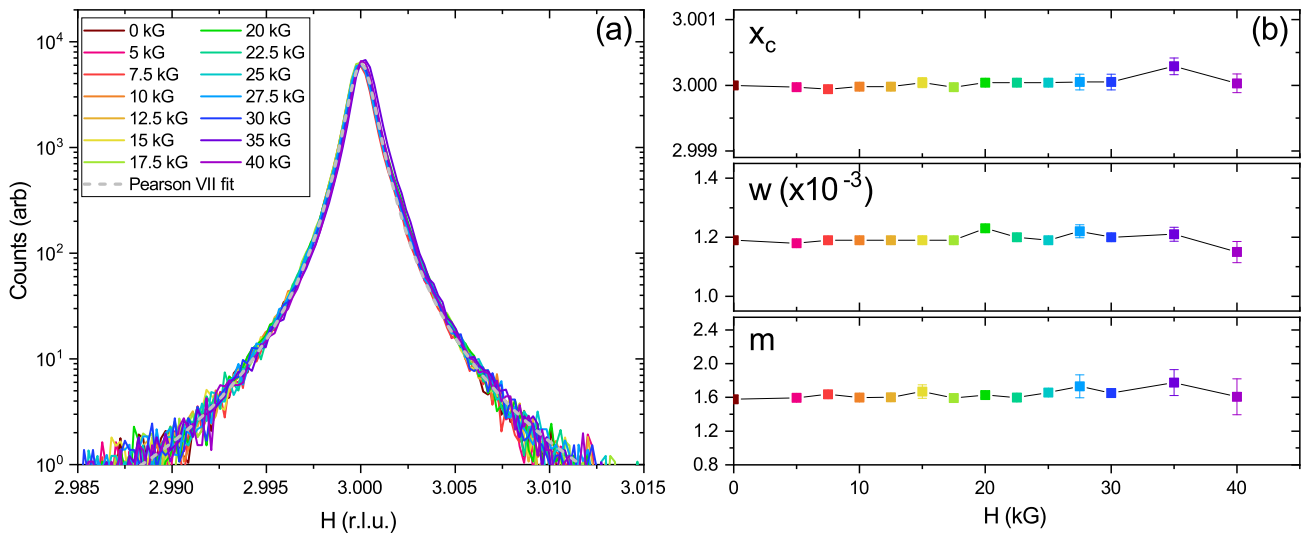}
    \caption{
(a) (300) peak in the $R\bar{3}m$ structure at 2.1 K as a function of applied field.
As field increases, the sample exits the superconducting state, but no change in the (300) peak is observed.
(b) Pearson-VII fit parameters $x_c$ (top), $w$ (middle), $m$ (bottom) as a function of field.
}
    \label{figInFieldParameters}
\end{figure*}

A narrow window in H around the (300) diffraction peak was then scanned at a fixed temperature of 2.1 K in increasing fields from 0 to 40 kG (Fig.~\ref{figInFieldParameters}(a)).
This field range guarantees that the peak is measured on either side of and across the superconducting transition.
A Pearson VII fit \cite{Hall-JAC} of the form
\begin{multline*}
	y = y_0 + A \frac{2\Gamma(m)\sqrt{2^{-m}-1}}{\sqrt{\pi}\Gamma(m-1/2)w} \\
	\times \left(1 + 4*\frac{2^{-m}-1}{w^2} \left(x - x_c \right)^2 \right)^{-m}
\end{multline*}
was used on each dataset.
The panels of Fig.~\ref{figInFieldParameters}(b) show the evolution of the central value of the peak $x_c$ (top), the width parameter $w$ (middle), and the exponent $m$ (bottom); for each, there is essentially no shift as the sample exits the superconducting state.
This suggests that there is no crystallographic shift or distortion associated with the superconducting transition.
The uncertainties of the fit parameters $x_c$, $w$, and $m$ are $4\times10^{-5}$, $6\times10^{-6}$ and $7\times10^{-2}$, respectively, slightly increasing with increasing applied field.
Importantly, the observed variations are random.
In particular, there is no uniform trend, for instance towards increased $x_c$ or $w$ with increasing magnetic field.
We therefore conclude that the data shown in Fig.~\ref{figInFieldParameters}(a) and Fig.~\ref{figInFieldParameters}(b) indicate that there is no in-plane crystallographic distortion at the level of $1\times10^{-5}$ associated with the superconducting transition.
These results further support the model that the large two-fold in-plane anisotropy of superconducting properties of Sr$_x$Bi$_2$Sr$_3$ is electronic in origin, namely caused by nematic superconducting order parameter of $E_u$ symmetry.


\section{Conclusion}
Our experiments demonstrate the viability of sub-kelvin synchrotron diffraction measurements.
We estimate that the sample reached a temperature under 1 K with a full beam of approximately 10$^{11}~\gamma$/sec at an energy of 19.9 keV, and a base temperature of approximately 0.68 K with no beam.
By monitoring the resistance, we demonstrated that the Sr$_{0.1}$Bi$_2$Se$_3$ sample remains superconducting in-beam, verifying that our base temperature diffraction measurements were performed deep in the superconducting state.
We saw no difference between the high symmetry (300) peaks at the base temperature of $\sim$0.25$T_c$ and at $\sim$1.5$T_c$, and we saw no change in the (300) peak as the sample was driven into the normal state with an applied magnetic field.
Our results indicate that there is no in-plane crystallographic distortion of the average lattice structure at the level of $1\times10^{-5}$ associated with the superconducting transition.
While our XRD measurements do not reach the resolution of 10$^{-7}$ reported in recent low-temperature dilatometry experiments \cite{NBS-Lortz}, our results further support the model in which the large two-fold in-plane anisotropy of superconducting properties of Sr$_x$Bi$_2$Se$_3$ is not structural in origin but electronic, namely it is caused by a nematic superconducting order parameter of $E_u$ symmetry.
The multimodal measurement capability has proven essential for quantifying beam-induced sample heating which, in the present case, amounted to about 0.25 K.

\section*{Acknowledgements}
Work at Argonne National Laboratory was supported by the U.S. Department of Energy, Office of Science, Basic Energy Sciences, Materials Sciences and Engineering Division.
This research used resources of the APS, a US DOE Scientific User Facility, operated for the DOE Office of Science by Argonne National Laboratory under Contract DE-AC02-06CH11357.
Work at Brookhaven was supported by the Office of Basic Energy Sciences, Division of Materials Sciences and Engineering, U.S. Department of Energy under Contract No. DE-SC00112704


\bibliographystyle{apsrev4-2}

\end{document}